\documentclass[submitting]{nst}
\usepackage{tikz}
\usepackage{graphicx}

\usetikzlibrary{decorations.pathmorphing}
\usepackage[T2A,T1]{fontenc}
\usepackage[russian,english]{babel}
\usepackage{slashed}
\usepackage{mathtools}

\usepackage{changes}
\usepackage{enumitem} 
\usepackage{listings}
\begin{document}
\title{Probing the Collision Geometry via Two-Photon Processes in Heavy-Ion Collisions }\thanks{Supported in part by the National Key Research and Development Program of China under Contract No. 2022YFA1604900 and the National Natural Science Foundation of China (NSFC) under Contract No. 12175223 and 12005220. W. Zha is supported by Anhui Provincial Natural Science Foundation No. 2208085J23, Youth Innovation Promotion Association of Chinese Academy of Sciences and the Chinese Academy of Sciences (CAS) under Grants No. YSBR-088.} 
\author{Jiaxuan Luo}
\affiliation{State Key Laboratory of Particle Detection and Electronics, University of Science and Technology of China, Hefei 230026, China}
\author{Xinbai Li}
\affiliation{State Key Laboratory of Particle Detection and Electronics, University of Science and Technology of China, Hefei 230026, China}
\author{Zebo Tang}
\affiliation{State Key Laboratory of Particle Detection and Electronics, University of Science and Technology of China, Hefei 230026, China}
\author{Xin Wu}
\affiliation{State Key Laboratory of Particle Detection and Electronics, University of Science and Technology of China, Hefei 230026, China}
\author{Shuai Yang}
\affiliation{State Key Laboratory of Nuclear Physics and Technology, Institute of Quantum Matter, South China Normal University, Guangzhou 510006, China}
\author{Wangmei Zha}
\email[Corresponding author, ]{Wangmei Zha Address: No. 96 Jinzhai Road, Hefei city, China Tel: +86 551 63607940  Email: first@ustc.edu.cn}
\affiliation{State Key Laboratory of Particle Detection and Electronics, University of Science and Technology of China, Hefei 230026, China}
\author{Zhan Zhang}
\affiliation{State Key Laboratory of Particle Detection and Electronics, University of Science and Technology of China, Hefei 230026, China}

\begin{abstract}
The initial collision geometry, including the reaction plane, is crucial for interpreting collective phenomena in relativistic heavy-ion collisions, yet it remains experimentally inaccessible through conventional measurements. Recent studies propose utilizing photon-induced processes as a direct probe, leveraging the complete linear polarization of emitted photons whose orientation strongly correlates with the collision geometry. In this work, we employ a QED-based approach to systematically investigate dilepton production via two-photon processes in heavy-ion collisions at RHIC and LHC energies and detector acceptances. Our calculations reveal that dilepton emission exhibits significant sensitivity to the initial collision geometry through both the azimuthal angles of their emission (defined by the relative momentum vector of the two leptons) and the overall momentum orientation of the dilepton pairs. These findings highlight the potential of two-photon-generated dileptons as a novel, polarization-driven probe to quantify the initial collision geometry and reduce uncertainties in characterizing quark-gluon plasma properties.
\end{abstract}

\maketitle
\section{Introduction}
The primary objective of relativistic heavy-ion collisions is to create and study quark-gluon plasma (QGP), the deconfined state of strongly interacting matter believed to exist microseconds after the Big Bang \cite{ref1_Braun-Munzinger:2007edi, ref2_STAR:2005gfr, Aoki:2006we}. Over decades of experimental and theoretical efforts, the formation of QGP in laboratory conditions has been firmly established, marking a pivotal achievement in high-energy nuclear physics. Current research focuses on characterizing QGP properties—such as its transport coefficients, equation of state, and response to extreme electromagnetic fields—and mapping the phase diagram of quantum chromodynamics (QCD) matter \cite{Ding:2015ona, Stoecker:2004qu, Zhang:2023ppo, Fukushima:2010bq}.

Central to these investigations are probes of collective phenomena in heavy-ion collisions, including anisotropic flow, global spin polarization, and chiral magnetic effects \cite{STAR:2004jwm, Kharzeev:2015znc, Liang:2004ph, STAR:2021mii, Anderle:2021wcy, Chen:2024aom}. These observables, measured extensively across collision energies and systems (e.g., by STAR, ALICE, and CMS collaborations), reveal that QGP behaves as a near-perfect fluid with remarkable vorticity, intense electromagnetic fields, and signatures of chiral symmetry restoration \cite{Bernhard:2019bmu, Becattini:2020ngo}. However, a critical challenge persists: the initial collision geometry (e.g., the reaction plane and participant eccentricity) cannot be directly accessed in experiments \cite{Miller:2007ri}. Current methods infer geometry indirectly via final-state momentum anisotropies, inherently conflating initial-state properties with medium-induced effects, non-flow correlations, and event-by-event fluctuations \cite{Qin:2010pf, PHOBOS:2010ekr, Jia:2014jca}. This introduces systematic biases that obscure quantitative links between QGP properties and initial conditions, underscoring the need for direct probes of collision geometry.

Recent advances propose photon-induced processes in hadronic heavy-ion collisions (HHICs) as a novel pathway to access initial geometry \cite{Wu:2022exl}. When relativistic nuclei collide, their strong electromagnetic fields generate quasireal photons, with polarization vectors oriented perpendicular to the motion of the colliding nucleus \cite{Krauss:1997vr, Li:2019yzy, STAR:2019wlg}. These polarized photons initiate coherent processes through both QED and QCD mechanisms:
\begin{itemize}
    \item \textbf{QED-dominated channels}: Photon-photon fusion into dilepton pairs (e.g., $e^+e^-$ via $\gamma\gamma \to e^+e^-$), where the photon's linear polarization dictates angular correlations in the decay products \cite{STAR:2018ldd, ATLAS:2018pfw, Zha:2018ywo, Shao:2022cly}.
    \item \textbf{QCD-assisted processes}: Coherent photoproduction of vector mesons (e.g., $J/\psi$) through photon-Pomeron interactions, where the polarization transfers to the produced meson \cite{STAR:2019yox, Zha:2020cst, Shen:2024eeb, Wu:2024jqn}.
\end{itemize}
The photon polarization direction is geometrically encoded by the initial collision configuration, enabling polarization-based probes of the reaction plane. Pioneering studies have validated this concept: Xiao \textit{et al.}~\cite{Xiao:2020ddm} first predicted a quadrupole modulation in the azimuthal angle of dileptons (defined by the relative momentum direction $\vec{p}_1 - \vec{p}_2$) from $\gamma\gamma$ processes, directly reflecting the photon polarization anisotropy. Subsequently, Wu \textit{et al.}~\cite{Wu:2022exl} proposed leveraging vector meson photoproduction to reconstruct the reaction plane through polarization-induced decay asymmetries. Recently, STAR measurements of coherently photoproduced $J/\psi$ mesons~\cite{kaiyang:star2024hp} in HHICs confirmed alignment between the $J/\psi$ decay anisotropy and the reaction plane, cementing photon polarization as a direct probe of initial geometry.

In this work, we employ our established QED framework~\cite{Zha:2018tlq, Zha:2021jhf, Li:2023yjt} to quantify dilepton production in HHICs, focusing on their dual sensitivity to both azimuthal emission angles ($\Delta\phi \propto \vec{p}_1 - \vec{p}_2$), and momentum orientation ($\Phi_{\rm pair} \propto \vec{p}_1 + \vec{p}_2$) relative to the reaction plane. Unlike prior studies primarily utilizing $\Delta\phi$-dependent observables, we demonstrate for the first time—that the dilepton pair's total momentum orientation ($\Phi_{\rm pair}$) exhibits significantly enhanced sensitivity to geometric information. 
Our calculations at RHIC and LHC energies reveal that the geometric constraints from $\Phi_{\rm pair}$ exceed those of $\Delta\phi$-based analyses in precision, while both observables provide orthogonal perspectives on reaction plane, bypassing systematic uncertainties inherent to traditional flow analyses. This establishes the QED-calibrated dilepton framework—simultaneously leveraging $\Delta\phi$ and $\Phi_{\rm pair}$—as a multi-dimensional probe for model-independent extraction of initial geometry.

\section{Methodological Framework}
\label{sec:method}

The Equivalent Photon Approximation (EPA) provides a computationally efficient framework for calculating total cross sections in heavy-ion collisions through the convolution of photon fluxes with the elementary $\gamma\gamma \to \ell^+\ell^-$ Breit-Wheeler process \cite{Krauss:1997vr, Klein:2016yzr}:
\begin{equation}
\sigma_{\ell^+\ell^-} = \int \frac{d\omega_1}{\omega_1} \frac{d\omega_2}{\omega_2} n(\omega_1) n(\omega_2) \sigma_{\gamma\gamma}(\omega_1,\omega_2)
\end{equation}
where $n(\omega)$ represents the photon flux density and $\sigma_{\gamma\gamma}$ is the elementary Breit-Wheeler cross-section. While EPA accurately describes integrated cross sections, it becomes progressively inadequate for differential observables due to its inherent neglect of photon transverse momentum correlations and polarization interference effects between scalar ($\sigma_s$) and pseudoscalar ($\sigma_{ps}$) interaction channels.

To overcome these limitations, we employ a lowest-order QED formulation based on external field approximation. Following Ref.~\cite{Vidovic:1992ik}, the electromagnetic potentials of colliding nuclei in Lorentz gauge are:
\begin{equation}
\label{eq:potentials}
\begin{split}
A_1^\mu(q_1,b) &= -2\pi (Z_1 e) e^{iq_1^\tau b_\tau} \delta(q_1^\nu u_{1\nu}) \frac{F_1(-q_1^2)}{q_1^2} u_1^\mu \\
A_2^\mu(q_2,0) &= -2\pi (Z_2 e)  \delta(q_2^\nu u_{2\nu}) \frac{F_2(-q_2^2)}{q_2^2} u_2^\mu
\end{split}
\end{equation}
where $q_{1, 2}$ are the equivalent photon four-momenta, $u_{1,2} = \gamma(1,0,0,\pm v)$ are four-velocities in the collider frame, $b$ is the impact parameter of collision, and $F(-q^2)$ denotes the nuclear charge form factor obtained via Fourier transform of the charge density distribution. The charge density for a symmetrical nucleus is characterized by the Woods-Saxon profile:
\begin{equation}
\rho(r) = \frac{\rho_0}{1 + \exp\left[(r - R_{\text{WS}})/d\right]}
\label{eq:charge_density}
\end{equation}
where $R_{\text{WS}}$ represents the nuclear radius and $d$ denotes the surface diffuseness parameter, both determined from electron scattering measurements \cite{DeJager:1974liz, DeVries:1987atn}. The normalization constant $\rho_0$ ensures $\int_0^\infty \rho(r) 4\pi r^2 dr = A$.

With the direct and cross Feynman diagrams of the lowest- \pagebreak order two-photon interaction for lepton pair creation, the matrix element can be expressed as~\cite{Hencken:1994my}:
\begin{equation}
\mathcal{M} = \bar{u}(p_{-}) \hat{\mathcal{M}} v(p_{+})
\end{equation}
where
\begin{equation}
\begin{aligned}
\hat{\mathcal{M}} = & -i e^2 \int \frac{\mathrm{d}^4 q_1}{(2\pi)^4} \slashed{A}_{1}(q_1)
\frac{\slashed{p}_{-} - \slashed{q}_1 + m}{(p_{-} - q_1)^2 - m^2} \slashed{A}_{2}(q_2) \\
& -i e^2 \int \frac{\mathrm{d}^4 q_1}{(2\pi)^4} \slashed{A}_{2}(q_2)
\frac{\slashed{q}_1 - \slashed{p}_{+} + m}{(q_1 - p_{+})^2 - m^2} \slashed{A}_{1}(q_1) \\
= & -i\left(\frac{Z e^2}{2\pi}\right)^2 \frac{1}{2\beta} \int \mathrm{d}^2 q_{1\perp}
\frac{1}{q_1^2} \frac{1}{q_2^2} \exp\left(\mathrm{i} q_{1\perp} \mathbf{b}\right) \\
 \times & \left\{
\frac{\slashed{\mu}^{(1)}(\slashed{p}_{-} - \slashed{q}_1 + m)\slashed{\mu}^{(2)}}
{[(p_{-} - q_1)^2 - m^2]}
+ \frac{\slashed{\mu}^{(2)}(\slashed{q}_1 - \slashed{p}_{+} + m)\slashed{\mu}^{(1)}}
{[(q_1 - p_{+})^2 - m^2]}
\right\},
\end{aligned}
\end{equation}
with $q_2 \equiv p_{+} + p_{-} - q_1$. The four-momenta  of the produced leptons are denoted by  $p_{+}$ and $p_{-}$, while the longitudinal components of the quasi-real photon four-momenta $q_1$ are constrained through the relations:
\begin{gather}
q_1^0 = \frac{1}{2}\left[(\varepsilon_{+} + \varepsilon_{-}) + \beta(p_{+}^z + p_{-}^z)\right] \\
q_1^z = q_1^0/\beta
\end{gather}
where $\varepsilon_{\pm}$ are the lepton energies, $m$ is the lepton mass, and $\beta$ represents the relativistic collider velocity factor.
The probability of the lowest order pair production can then be straightforwardly expressed as:
\begin{equation}
\begin{aligned}
P(p_{+}, p_{-}; b) &= \sum_{\rm spin}|{\cal M}|^2 \\
= \frac{4}{\beta^2}  \int d^2 & q_{1 \perp}\,  d^2 \Delta q_{\perp}\,
\mathrm{e}^{\mathrm{i}\Delta q_{\perp}\cdot b} 
\prod_{i=0,1,3,4} \!\! \frac{F(N_i)}{N_i}\; \\
\times \operatorname{Tr} \Bigg\{
(\slashed{p}_{-} & + m) \bigg[
N_{2D}^{-1} \slashed{\mu}_1 (\slashed{p}_{-} - \slashed{q}_1 + m) \slashed{\mu}_2 \\
&\ \ \ \ \ \ \ \ \quad + N_{2X}^{-1} \slashed{\mu}_2 (\slashed{q}_1 - \slashed{p}_{+} + m) \slashed{\mu}_1 
\bigg] \\
\times ( & \slashed{p}_{+}  - m) \bigg[
N_{5D}^{-1} \slashed{\mu}_2 (\slashed{p}_{-} - \slashed{q}_1 + \slashed{\Delta q} + m) \slashed{\mu}_1 \\
&\ \ \quad + N_{5X}^{-1} \slashed{\mu}_1 (\slashed{q}_1 + \slashed{\Delta q} - \slashed{p}_{+} + m) \slashed{\mu}_2 
\bigg] \Bigg\},
\end{aligned}
\label{eq:production_probability}
\end{equation}
with
\begin{equation}
\begin{aligned}
&N_0 = -q_1^2, \\ 
&N_1 = -[q_1 - (p_{+} + p_{-})]^2, \\
&N_3 = -(q_1 + \Delta q)^2, \\ 
&N_4 = -[\Delta q + (q_1 - p_{+} - p_{-})]^2, \\
&N_{2D} = -(q_1 - p_{-})^2 + m^2, \\ 
&N_{2X} = -(q_1 - p_{+})^2 + m^2, \\
&N_{5D} = -(q_1 + \Delta q - p_{-})^2 + m^2, \\
&N_{5X} = -(q_1 + \Delta q - p_{+})^2 + m^2,
\end{aligned}
\label{eq:denominators}
\end{equation}
where $F(N_{i})$ represents the photon propagator \cite{Lee:1999eya, Lee:2001nt} and $\Delta q$ is the four-momentum difference between $q_1$ and $q'_1$ ($\Delta q \equiv q_1 - q'_1$), with propagator assignments organized as:
\begin{itemize}
\item\ \textbf{$\mathbf{\textit{i}=0,3}$ terms}: originate from nucleus 1 (direct/conjugate photons $q_1$ and $q'_1$);
\item\ \textbf{$\mathbf{\textit{i}=1,4}$ terms}: sourced from nucleus 2 (direct/conjugate photons $q_2$ and $q'_2$).
\end{itemize}

\begin{table*}
\caption{Fiducial cuts implemented in the calculation.}
\centering
\begin{tabular}{lc|ccccc}
\toprule
\multicolumn{2}{c|}{Process and beam energy} & $p_{Tl}$ (GeV/c)& $\eta_{l}$ & $P_{Tll}$ (GeV/c) &$Y_{ll}$ & $M_{ll}$ (GeV/${\text c^{2}}$)\\
\midrule
$\gamma \gamma \rightarrow e^+ e^-(\mu^+ \mu^-)$ &Au+Au $\sqrt{s_{\rm NN}}=200 $ GeV & $(0.2,+\infty)$& $(-1.0,1.0)$& $(0,0.3)$& $(-1.0,1.0)$& $(0.4,2.6)$\\
$\gamma \gamma \rightarrow e^+ e^-$ &Pb+Pb $\sqrt{s_{\rm NN}}=5.02 $ TeV& $(0.5,+\infty)$& $(-1.0,1.0)$& $(0,0.3)$&  $(-1.0,1.0)$ &$(1.0,2.8)$\\
$\gamma \gamma \rightarrow \mu^+ \mu^- $ &Pb+Pb $\sqrt{s_{\rm NN}}=5.02$ TeV & $(4.0,+\infty)$& $(-2.4,2.4)$& $(0,0.3)$& $(-2.4,2.4)$ & $(8.0,100.0)$\\
\bottomrule
\end{tabular}
\label{table1}
\end{table*}
We construct angular correlations through: 
\begin{align}
\Phi_{\rm{pair}} &\equiv \phi_{\vec{p}_+ + \vec{p}_-} - \phi_b \\
\Delta\phi &\equiv \phi_{\vec{p}_+ - \vec{p}_-} - \phi_b
\end{align}
where $\phi_b$ denotes the azimuthal angle of impact parameter $\vec{b}$. The differential distribution $dN/d\Phi_{\rm{pair}}(d\Delta\phi)$ is obtained by integrating over:
\begin{equation}
\int_{\Omega_{\rm phase}} \!\!\! \delta(\Phi_{\rm pair}(\Delta \phi)-f[\vec{p}_\pm,\vec{b}]) \, P(p_+,p_-;b) \, \mathcal{J} \, d\Omega_{\rm reduce}
\end{equation}
where $f[\vec{p}_\pm,\vec{b}]$ maps momentum combinations to angular differences, $\mathcal{J}$ contains Jacobian factors from coordinate transformations, and $d\Omega_{\rm reduce}$ denotes integration over non-angular momentum components.
The high-dimensional phase space integration is performed using the \textsc{Vegas} adaptive Monte Carlo algorithm \cite{Lepage:1977sw} within the ROOT framework \cite{Antcheva:2009zz}.

\begin{figure}[t]
\centering
\includegraphics[width=0.4\textwidth]{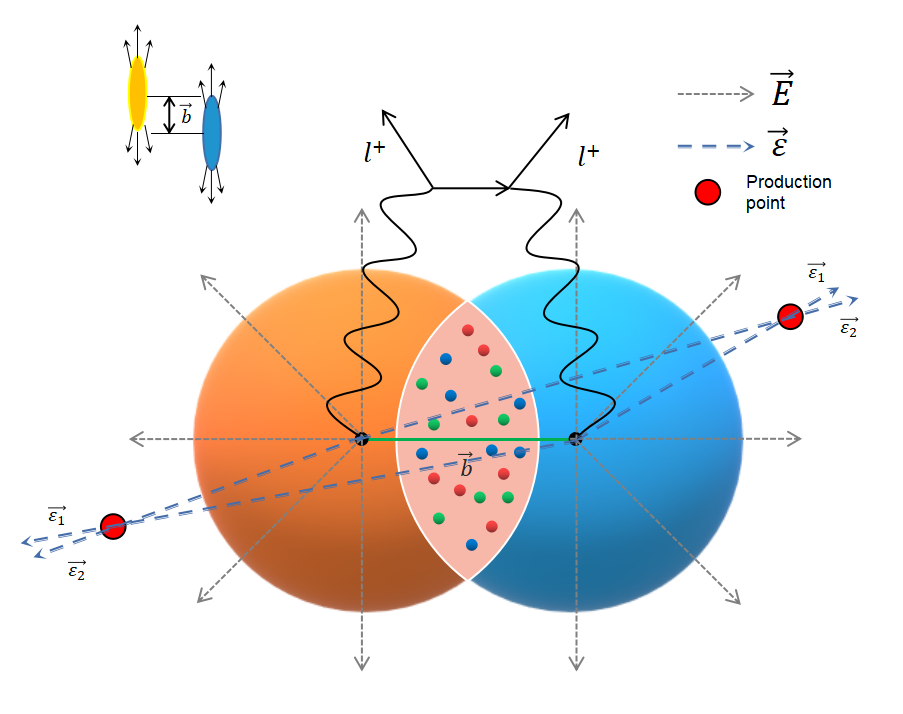}
\caption{Electromagnetic field configuration in transverse plane with collision impact parameter $\vec{b}$. }
\label{fig:dilepton_schematic}
\end{figure}

\begin{figure}[ht]
  \centering
  \begin{minipage}{0.4\textwidth}
    \includegraphics[width=\textwidth]{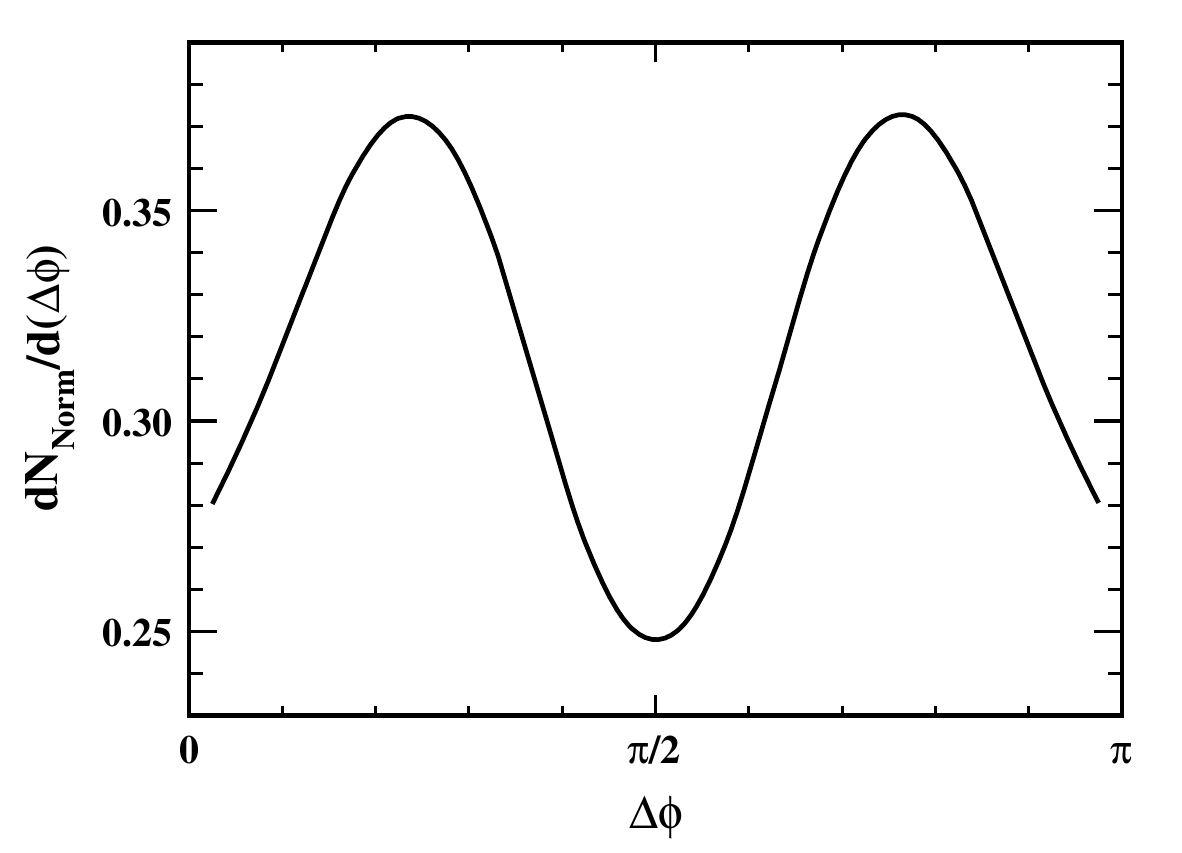}
  \end{minipage}
  \begin{minipage}{0.4\textwidth}
    \includegraphics[width=\textwidth]{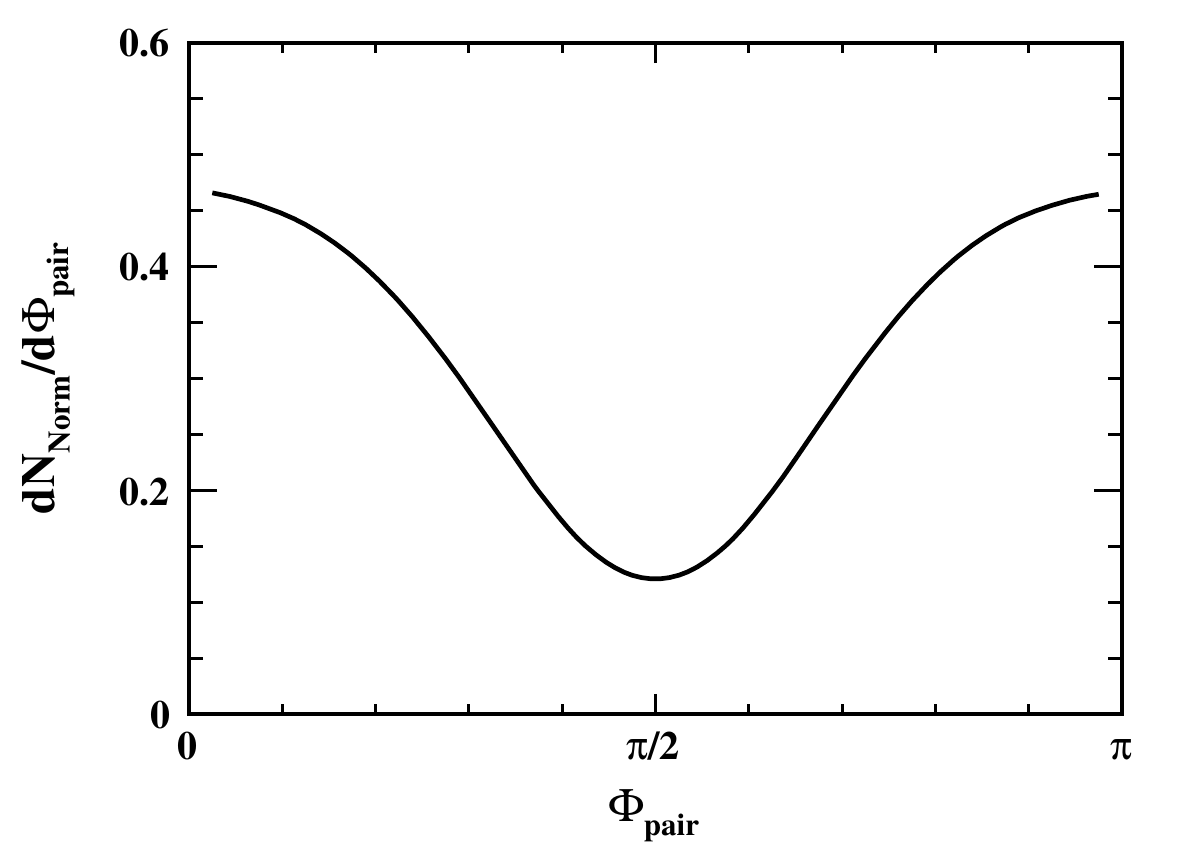}
  \end{minipage}
   \caption{The periodic oscillation pattern in the alignment of the $\phi_{\vec{p}_+ + \vec{p}_-}$ and $\phi_{\vec{p}_+ - \vec{p}_-}$ related to the reaction plane $\phi_b$ for the lepton pair production in Au+Au peripheral collisions at $\sqrt{s_{\rm NN}} = 200$~GeV with impact parameter $b=11$ fm.}
  \label{fig:2}
\end{figure}

\begin{figure}[ht]
  \centering
  \includegraphics[width=0.48\textwidth]{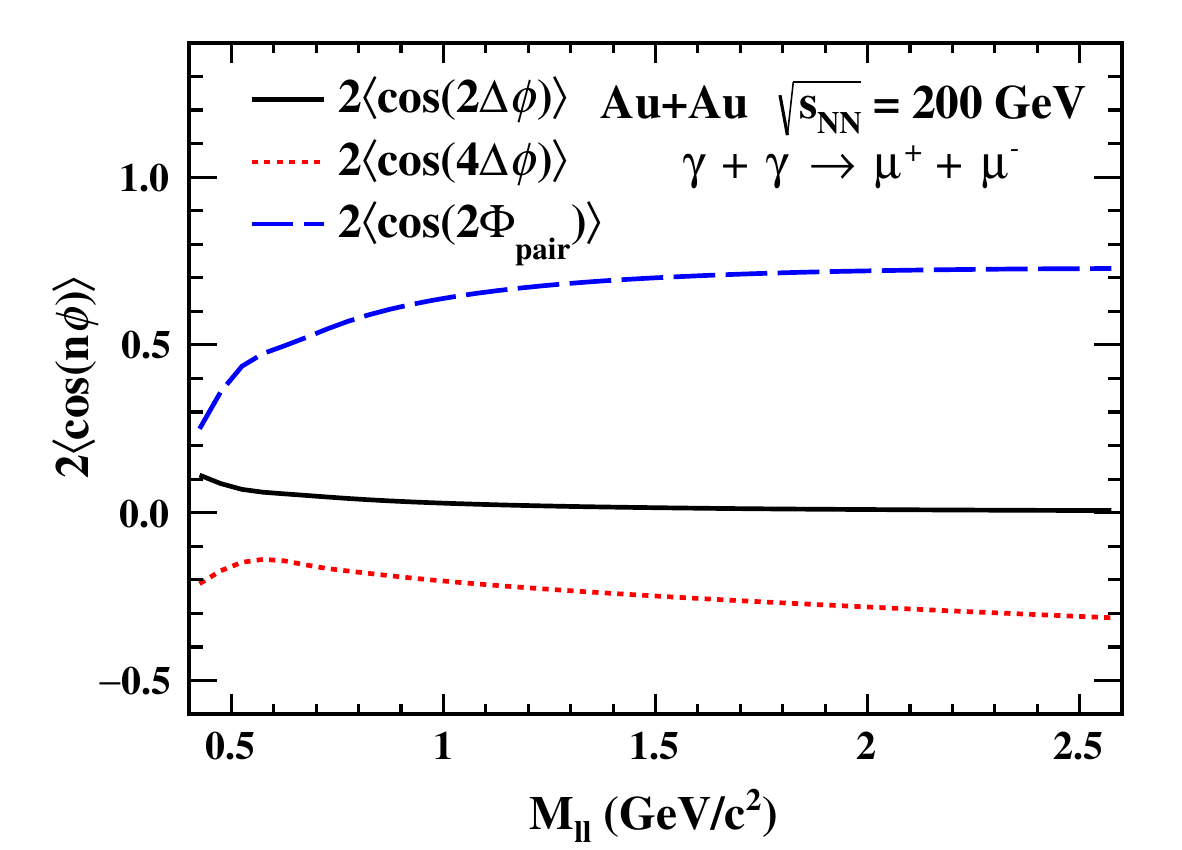}
   \caption{Second- and fourth-order modulation coefficients 2$\langle \rm{cos(n\phi)} \rangle$ for $\Delta\phi$ and $\Phi_{\rm{pair}}$ as a function of dimuon invariant mass in Au+Au peripheral collisions at $\sqrt{s_{\rm NN}} = 200$~GeV with impact parameter $b=11$ fm.}
  \label{fig:3}
\end{figure}

\begin{figure}[ht]
  \centering
  \begin{minipage}{0.48\textwidth}
    \includegraphics[width=\textwidth]{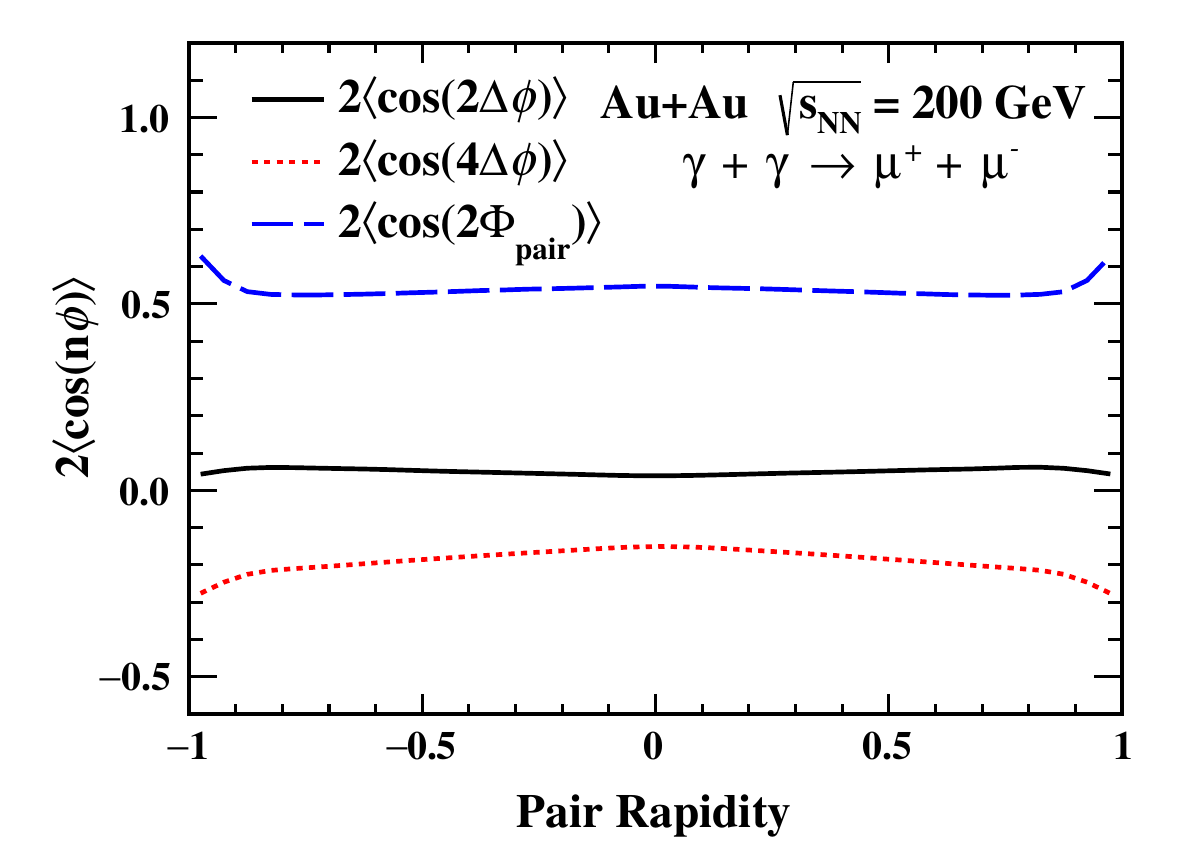}
  \end{minipage}
   \caption{Second- and fourth-order modulation coefficients 2$\langle \rm{cos(n\phi)} \rangle$ for $\Delta\phi$ and $\Phi_{\rm{pair}}$ as a function of dimuon pair rapidity in Au+Au peripheral collisions at $\sqrt{s_{\rm NN}} = 200$~GeV with impact parameter $b=11$ fm.}
  \label{fig:4}
\end{figure}

\begin{figure*}[!hbtp]
  \centering
\includegraphics[width=0.48\linewidth]{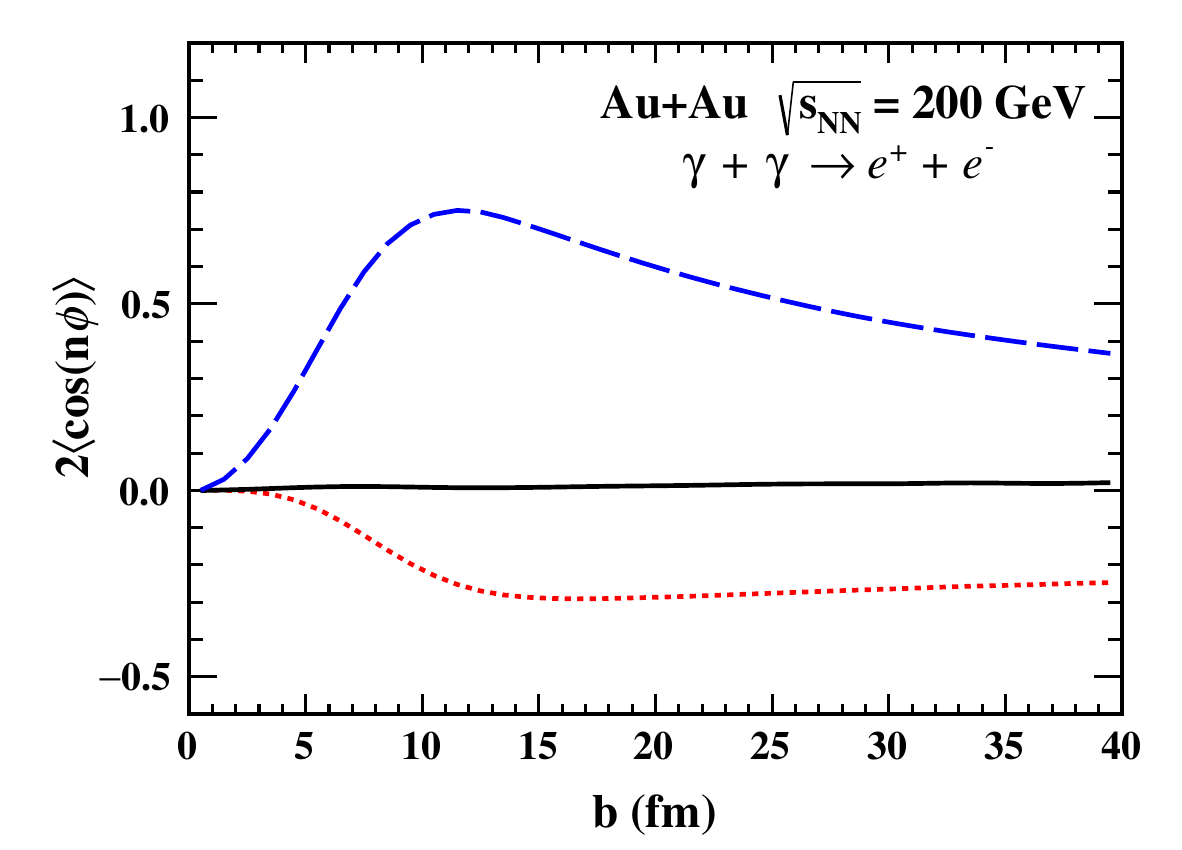}
\includegraphics[width=0.48\linewidth]{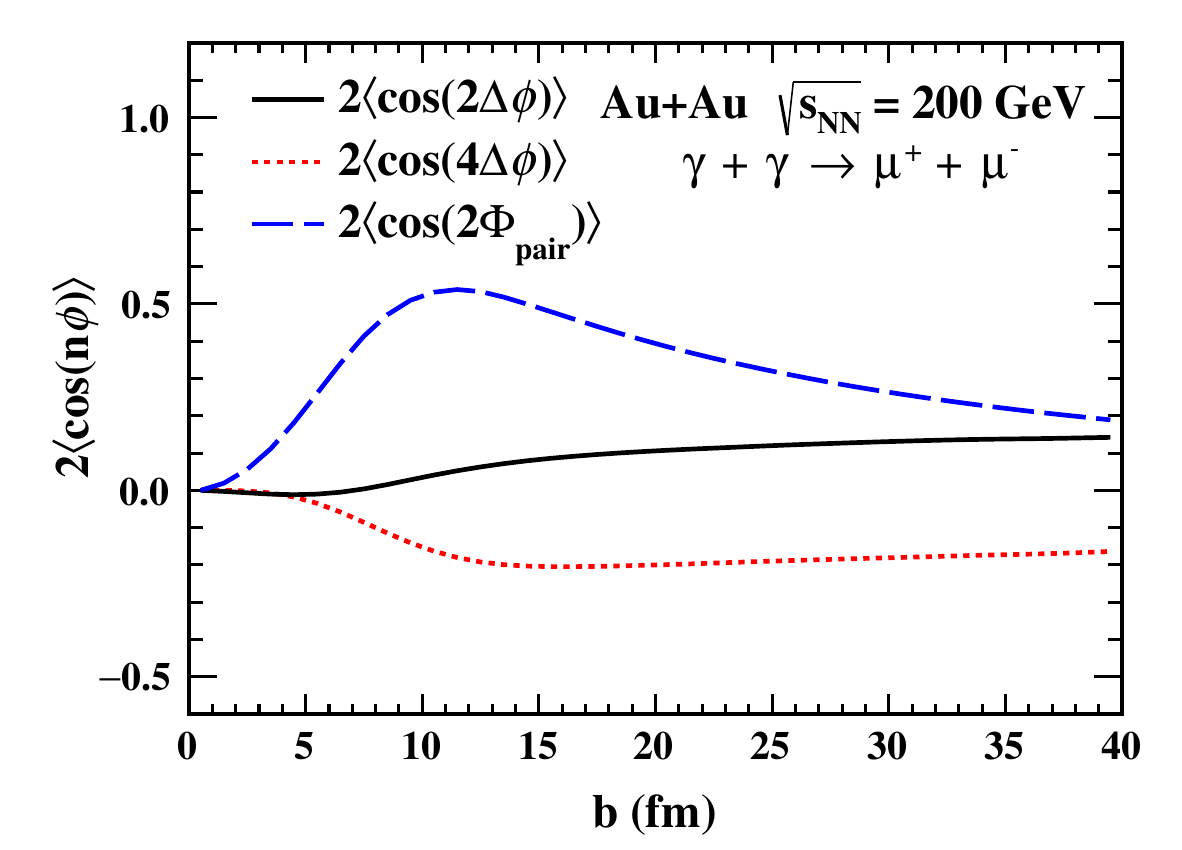}
\includegraphics[width=0.48\linewidth]{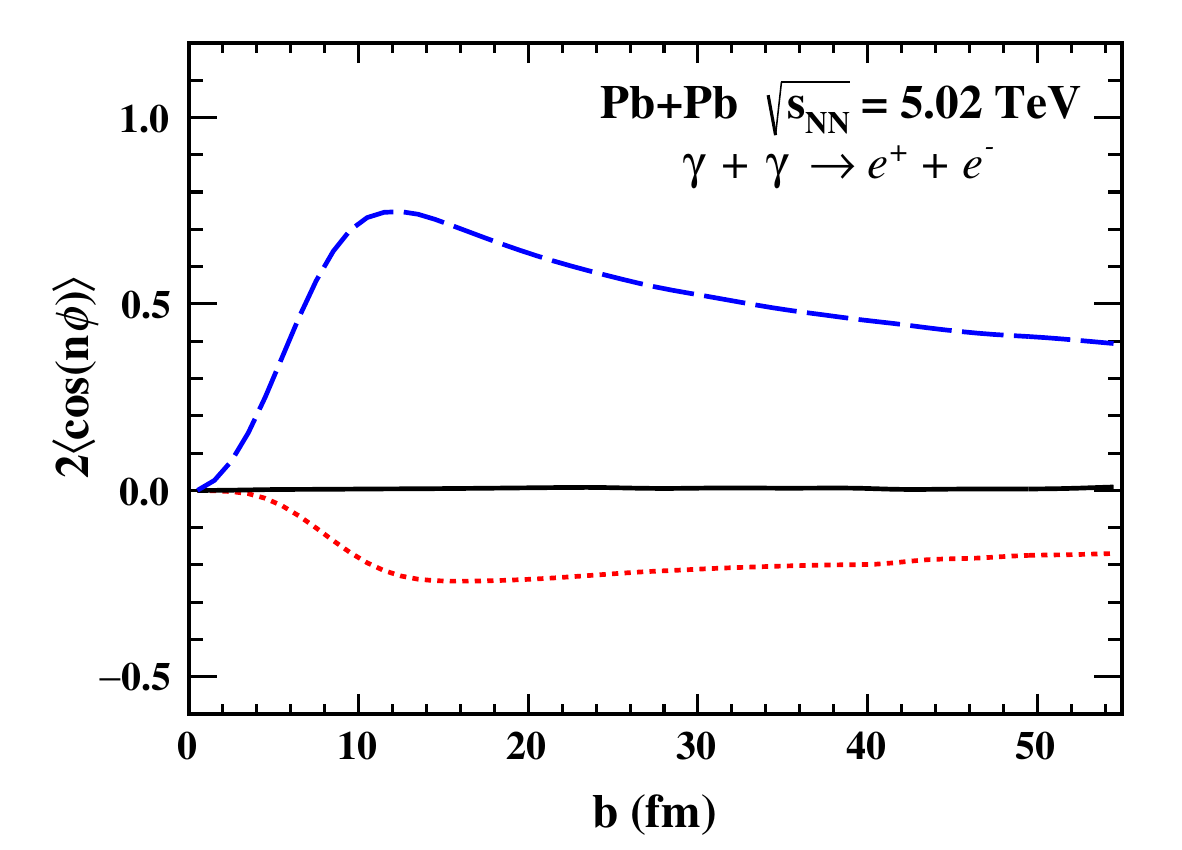}
\includegraphics[width=0.48\linewidth]{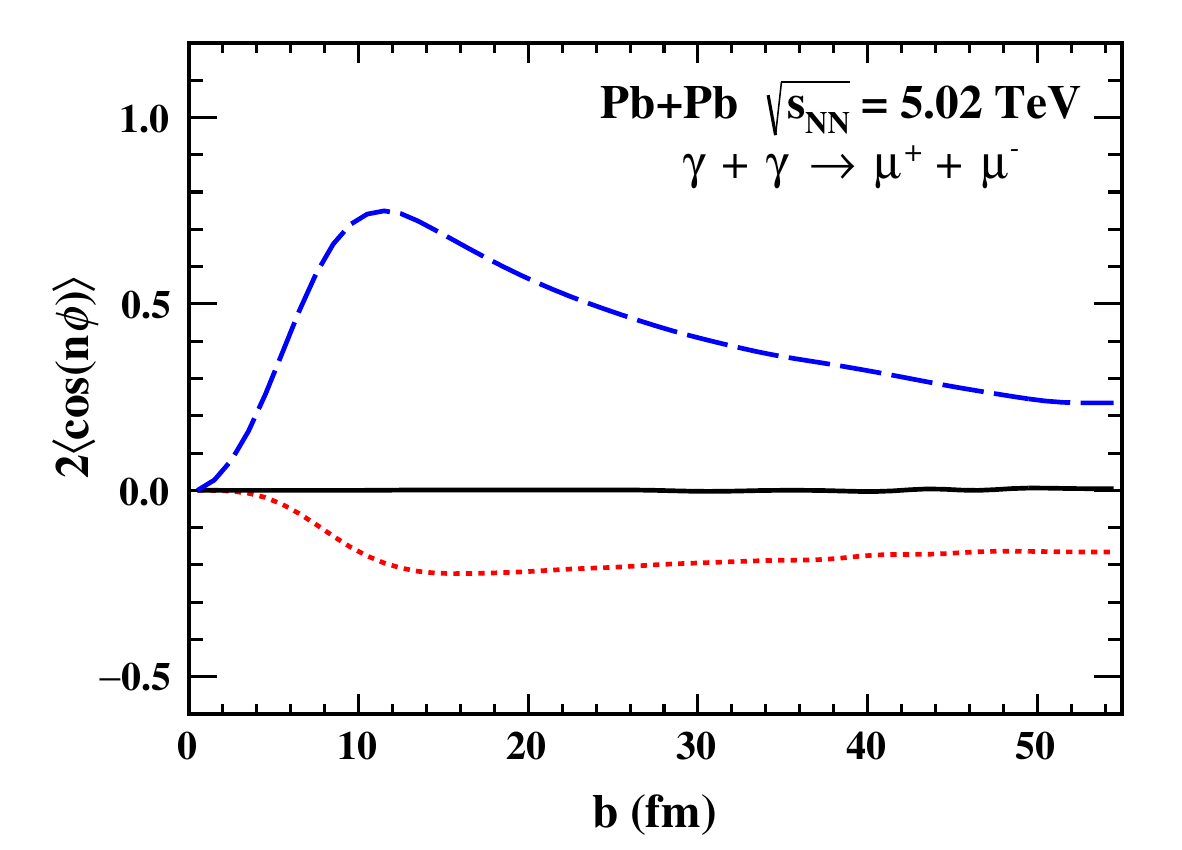}
\caption{Second- and fourth-order modulation coefficients 2$\langle \rm{cos(n\phi)} \rangle$ for $\Delta\phi$ and $\Phi_{\rm{pair}}$ as a function of impact parameter in Au+Au collisions at $\sqrt{s_{\rm NN}} = 200$~GeV (RHIC) and Pb+Pb collisions at $\sqrt{s_{\rm NN}} = 5.02$~TeV (LHC).}
\label{fig:5}
\end{figure*}

Although the QED formulation  provides precise calculations of angular correlations, the dynamical mechanism through which photon polarization induces angular asymmetry (with respect to impact parameter) remains implicit. To unveil the physical origins of this connection, we trace how the field geometry (governing polarization directions) dynamically "locks" onto momentum-space characteristics of the produced particles. Consider the electromagnetic field configuration in peripheral heavy-ion collisions: the quasi-real photons carry linear polarization determined by the classical electromagnetic fields.  Specifically, the electric field vector \(\vec{\varepsilon}\) at a transverse position \(\vec{r}_\perp\) (defined with respect to the nucleus center as the origin) satisfies the radial constraint:
\[
\vec{\varepsilon}(\vec{r}_\perp) \parallel \vec{r}_\perp,
\]
indicating that the electric field direction is radially oriented away from the nucleus center, as illustrated by the field line distribution in Fig.~\ref{fig:dilepton_schematic}.
This spatial polarization pattern becomes imprinted on the photon momentum distribution through the Fourier relation $\vec{k}_\perp \leftrightarrow \vec{r}_{\perp}$, leading to the momentum-polarization locking
\begin{equation}
    \vec{\varepsilon}(\vec{k}_\perp) \parallel \vec{k}_\perp\,.
    \label{eq:momentum_polarization}
\end{equation}

In the dilepton production process $\gamma\gamma \to \ell^+\ell^-$, the scattering amplitude depends critically on the relative orientation of the photon polarizations:
\begin{equation}
    \mathcal{M} \propto \left( \vec{\varepsilon}_1 \cdot \vec{\varepsilon}_2 \right) \times \text{lepton current}.
    \label{eq:amplitude}
\end{equation}
The polarization overlap $|\vec{\varepsilon}_1 \cdot \vec{\varepsilon}_2|$ reaches maximum when the two photons' transverse momenta are collinear ($\vec{k}_{\perp 1} \parallel \vec{k}_{\perp 2}$). As illustrated in Fig.~\ref{fig:dilepton_schematic}, this collinear condition forces the pair orientation $\Phi_{\rm pair}$ to align with the impact parameter. The observed $\Delta\phi$ asymmetry stems from the geometric suppression of polarization correlations due to angular misalignment. Specifically, as demonstrated in Fig.~\ref{fig:dilepton_schematic}, when the transverse momenta $\vec{k}_{\perp 1}$ and $\vec{k}_{\perp 2}$ acquire non-collinear components (induced by a finite impact parameter $\vec{b}$), their polarization vectors $\vec{\varepsilon}_1$ and $\vec{\varepsilon}_2$ become mismatched. These misalignments introduce a characteristic modulation in the angular distribution, dominated by $\cos(2\Delta\phi)$ and $\cos(4\Delta\phi)$ terms with respect to the impact parameter axis — direct consequences of the spin-1 nature of photons and spin-1/2 nature of leptons in the $\gamma\gamma \to \text{dilepton}$ production process \cite{STAR:2019wlg}.


\section{Results}
\label{sec:results}

Figure~\ref{fig:2} illustrates the azimuthal modulations of dilepton pairs with respect to the reaction plane in peripheral Au+Au collisions at $\sqrt{s_{\rm NN}} = 200$ GeV (RHIC) with impact parameter $b = 11$ fm. The detailed fiducial cuts applied in the calculation are presented in Table~\ref{table1}. The $\Delta\phi$ distribution exhibits a dominant $\cos(4\Delta\phi)$ pattern due to the spin of the interacting photons. A secondary $\cos(2\Delta\phi)$ modulation, originating from lepton mass effects, becomes visible as the finite muon mass breaks the helicity conservation. In contrast, the $\Phi_{\rm pair}$ distribution shows a pronounced $\cos(2\Phi_{\rm pair})$ modulation, directly reflecting the alignment of the dilepton total momentum with the reaction plane. This alignment arises from the momentum-polarization locking effect discussed in Section~\ref{sec:method}, where the global polarization geometry encoded in the electromagnetic fields is imprinted onto the angular correlation of dilepton pairs.

The invariant mass dependence of anisotropic coefficients $\langle\cos(4\Delta\phi)\rangle$ and $\langle\cos(2\Phi_{\rm pair})\rangle$ is shown in Fig.~\ref{fig:3}. At low mass ($M_{\mu\mu} < 0.8$ GeV/$\text c^{2}$), the $\langle\cos(2\Delta\phi)\rangle$ component is visible due to the violation of helicity conservation. This coefficient diminishes as $M_{\mu\mu}$ increases beyond 0.8 GeV/$\text c^{2}$, suppressed by restoration of helicity conservation. The $\langle\cos(4\Delta\phi)\rangle$ coefficient displays a non-monotonic trend, initially decreasing to a local minimum at $M_{\mu\mu} \sim 0.8$ GeV/$\text c^{2}$ before rising again at higher mass. This behavior reflects competing contributions from two physical mechanisms: geometric depolarization effects dominate at intermediate mass, while polarization coherence strengthens at larger $M_{\mu\mu}$ as dileptons are preferentially produced near the nuclear periphery. In contrast, the $\langle\cos(2\Phi_{\rm pair})\rangle$ coefficient retains a consistently positive non-zero value and increases monotonically with dimuon invariant mass. This distinct mass dependence highlights its enhanced sensitivity as a global geometric observable to local kinematic configurations compared to $\langle\cos(4\Delta\phi)\rangle$.

Figure~\ref{fig:4} explores the rapidity dependence of these coefficients for $0.4 < M_{\mu\mu} < 2.6$ GeV/${\rm c}^2$. The magnitude of  $\langle\cos(4\Delta\phi)\rangle$ increases with $|y|$, driven by enhanced polarization alignment in forward/backward regions where photoproduction occurs closer to the nuclear surfaces. This spatial localization amplifies the correlation between photon polarization vectors and the impact parameter orientation. However, within the rapidity range $|y| < 0.8$, the $\langle\cos(2\Phi_{\rm pair})\rangle$ coefficient remains approximately constant and exhibits negligible rapidity dependence. This distinct behavior arises because the total momentum orientation of dilepton pairs depends primarily on the vector sum of photon momenta—a global property governed by the reaction plane geometry rather than local rapidity-dependent dynamics.

The impact parameter dependence of the modulation coefficients, calculated for both RHIC and LHC collision systems, is presented in Fig.~\ref{fig:5}. Central collisions ($b \to 0$) yield vanishing anisotropies as the polarization geometry becomes azimuthally symmetric. The magnitudes of the coefficients grow with increasing $b$, reaching maxima near $b \sim 2R_A$ ($R_A$ being the nuclear radius). Beyond this point, $\langle\cos(4\Delta\phi)\rangle$ and $\langle\cos(2\Phi_{\rm pair})\rangle$ stabilize and slightly decrease due to the depletion of photon flux coherence at ultra-peripheral separations. Remarkably, the $\langle\cos(2\Phi_{\rm pair})\rangle$ values consistently  exceed those of $\langle\cos(4\Delta\phi)\rangle$ across the entire $b$ range, demonstrating the superior sensitivity of pair momentum orientation to collision geometry. This systematic trend persists in Pb+Pb collisions at $\sqrt{s_{\rm NN}} = 5.02$ TeV (LHC), confirming the universality of QED-dominated polarization effects across collision energies.

Collectively, these results establish $\Phi_{\rm pair}$ correlations as a precision tool for initial geometry determination. The distinct dependence of $\langle\cos(2\Phi_{\rm pair})\rangle$ on mass and rapidity can effectively  mitigate the systematic bias induced by final-state interactions, while its strong $b$-dependence enables direct mapping between observable angular modulations and collision geometry. Complementary information from $\Delta\phi$ harmonics provides cross-checks on polarization purity—crucial for separating two-photon processes from competing hadronic backgrounds. The combined analysis of both observables opens a new pathway for model-independent extraction of the reaction plane in heavy-ion collisions, free from assumptions about medium evolution or hadronization dynamics.
\section{Summary}
\label{sec:summary}
In summary, we have employed a QED-based approach to systematically investigate how photon-induced processes can constrain initial collision geometry in heavy-ion collisions at RHIC and LHC energies.  
This work introduces a novel observable, the total momentum orientation of dilepton pair $\Phi_{\rm pair}$, which demonstrates significantly enhanced sensitivity to geometric features, exceeding $\Delta\phi$-based analyses in precision.
By combining the complementary sensitivities of photo-produced dilepton to primordial geometry via both $\Delta\phi$ and $\Phi_{\rm pair}$, our calculations establish the QED-calibrated dilepton framework as a robust, multi-dimensional probe that enables model-independent reconstruction of collision geometry, offers orthogonal constraints on reaction plane determination, and bypasses systematic uncertainties inherent to traditional flow methods. 
Future measurements of these observables are expected to provide direct and unambiguous experimental constraints on the primordial geometry in relativistic heavy-ion collisions.




\end{document}